\documentclass[a4paper,showpacs,preprintnumbers,amsmath,amssymb,superscriptaddress,floatfix]{jpconf}
\usepackage{graphicx}
\usepackage{amsmath}
\usepackage{bm}
\usepackage{mathrsfs}
\usepackage{color}
\usepackage{slashed}

\newcommand{\bald}[1]{{\bf #1}}

\newcommand{\eqf}[1]{\begin{equation}\begin{split}#1\end{split}\end{equation}}

\begin{document}

\title{Jets associated with $Z^0$ boson production in heavy-ion collisions at the LHC}

\author{R. B. Neufeld \\
{\it Los Alamos National Laboratory, Theoretical Division, MS B238, Los Alamos, NM 87545, U.S.A.}}

\date{\today}

\begin{abstract}
The heavy ion program at the LHC will present unprecedented opportunities to probe hot QCD matter, that is, the quark gluon plasma (QGP). Among these exciting new probes are high energy partons associated with the production of a $Z^0$ boson, or {\it $Z^0$ tagged jets}.  Once produced, $Z^0$ bosons are essentially unaffected by the strongly interacting medium produced in heavy-ion collisions, and therefore provide a powerful signal of the initial partonic energy and subsequent medium induced partonic energy loss.  When compared with theory, experimental measurements of $Z^0$ tagged jets will help quantify the jet quenching properties of the QGP and discriminate between different partonic energy loss formalisms.  In what follows, I discuss the advantages of tagged jets over leading particles, and present preliminary results of the production and suppression of $Z^0$ tagged jets in relativistic heavy-ion collisions at LHC energies using the Guylassy-Levai-Vitev (GLV) partonic energy loss formalism.
\end{abstract}

%\maketitle

The suppression of high energy partons in the quark gluon plasma (QGP), that is {\it jet quenching}, is one of the most striking results from the heavy-ion program at RHIC \cite{Gyulassy:2003mc,Jacobs:2004qv}.  Much theoretical effort has been invested in understanding the mechanism of this suppression and in particular of partonic energy loss \cite{jetquenching,vitev}.  Because of the limited center-of-mass energies available at RHIC, experimental measurements have thus far focused on leading particle suppression relative to $p+p$ collisions \cite{Adler:2006hu}.  Unfortunately, leading particle suppression alone is not sufficient to discriminate between partonic energy loss formalisms or to extract quantitatively the jet quenching properties of the QGP \cite{Bass:2008rv}.

\begin{figure}
\centerline{
\includegraphics[width = 0.5\linewidth]{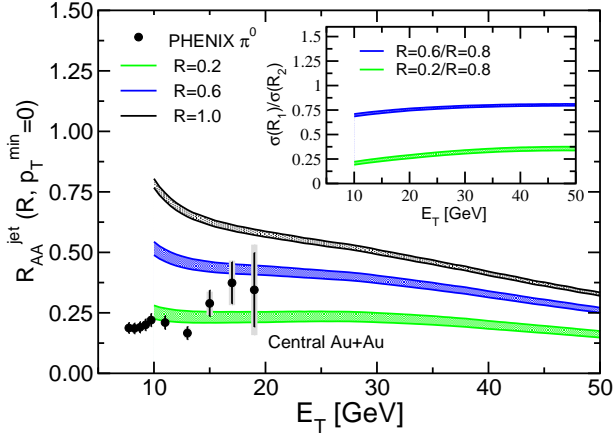}
}
\caption{\small{The figure, taken from Ref. \cite{Vitev:2009rd}, shows predictions for jet $R_{AA}$ as a function of Lorentz-invariant jet reconstruction cone size $R = \sqrt{(\Delta \phi)^2 + (\Delta y)^2}$ using the Guylassy-Levai-Vitev \cite{vitev} (GLV) partonic energy loss formalism.  The plot was done for impact parameter b = 3 fm in Au+Au collisions at $\sqrt{s} = 200$ GeV per nucleon.  The dependence of the suppression on cone size $R$ directly reflects the angular distribution of medium-induced radiation.  The insert shows the ratios of jet cross sections for selected values of $R$.}
}
\label{dingy}
\end{figure}

In order to constrain the underlying QCD theory of jet quenching, new and more differential observables are needed.  In particular, jet shapes \cite{Vitev:2008rz,Vitev:2009rd} and jets tagged with electromagnetic \cite{Srivastava:2002kg} or weakly interacting probes are especially promising. Jet reconstruction and analysis of the subsequent jet shapes provide an experimental handle on the medium-induced bremsstrahlung spectrum.  This is demonstrated in Figure \ref{dingy}, which is taken from Ref. \cite{Vitev:2009rd}.  The Figure shows predictions for the sensitivity of jet $R_{AA}$ on the Lorentz-invariant jet reconstruction cone size $R = \sqrt{(\Delta \phi)^2 + (\Delta y)^2}$ using the Guylassy-Levai-Vitev \cite {vitev} (GLV) partonic energy loss formalism.  The plot was done for impact parameter b = 3 fm in Au+Au collisions at $\sqrt{s} = 200$ GeV per nucleon.  The dependence of the suppression on cone size $R$ directly reflects the angular distribution of medium-induced radiation.  The heavy-ion program at the LHC will enable a high statistics experimental measurement of this feature for the first time.

In addition to jet shapes, {\it tagged jets} provide an exciting opportunity to quantify the jet quenching properties of the QGP.  Tagged jets refer to high energy partons which are produced in association with the tagging particle.  Electromagnetic and weakly interacting particles are ideal tags to study partonic energy loss because, once produced, they have negligible medium induced modifications.  Consequently, the tagging particle serves as a signal for the initial associated jet energy.  By reconstructing the tagged jet, one can obtain the amount and distribution of partonic energy loss.  The leptonic final states of the $Z^0$ boson are an especially attractive jet tag because the large invariant mass of the $Z^0$ boson makes it easy to distinguish from the background generated in a heavy-ion collision.  Again, the heavy-ion program at the LHC will enable a high statistics experimental measurement of this feature for the first time.

\begin{figure*}
\centerline{
\includegraphics[width = 0.4\linewidth]{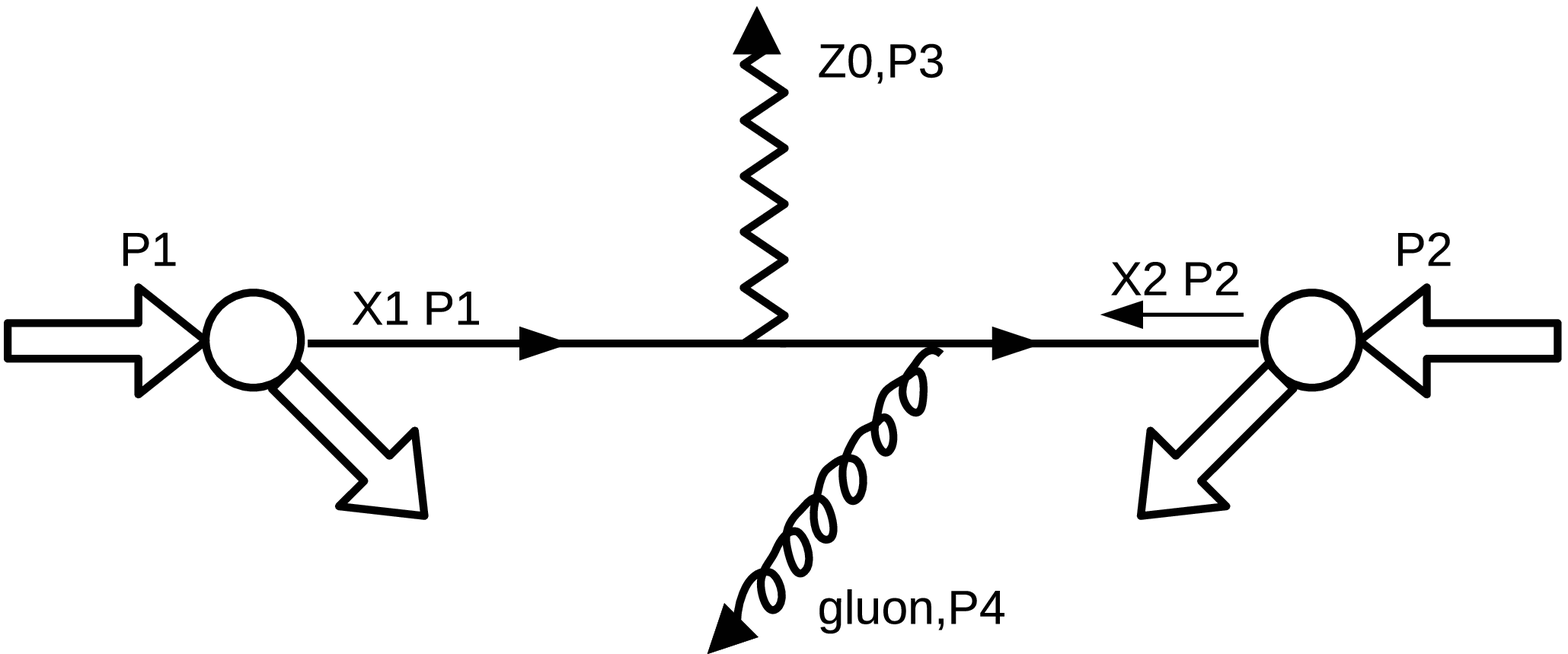}
\includegraphics[width = 0.4\linewidth]{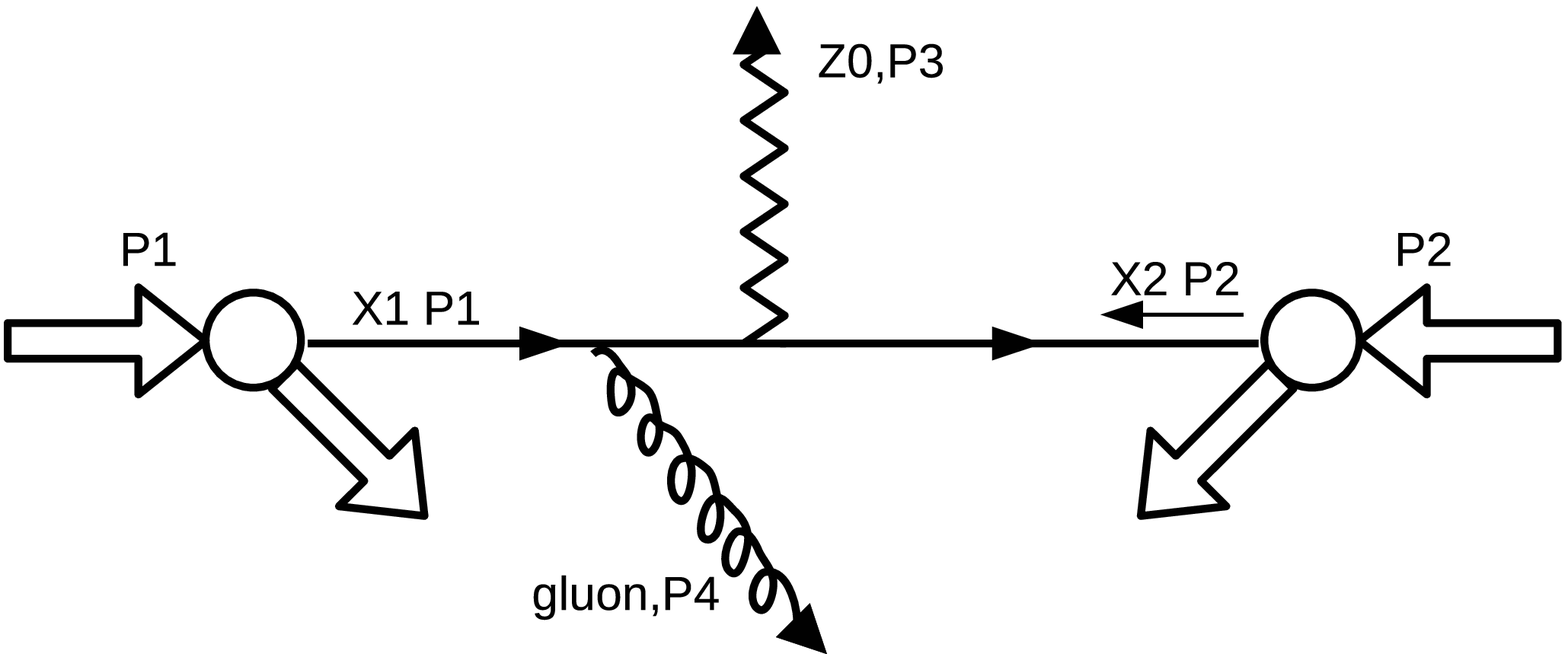}
}
\centerline{
\includegraphics[width = 0.4\linewidth]{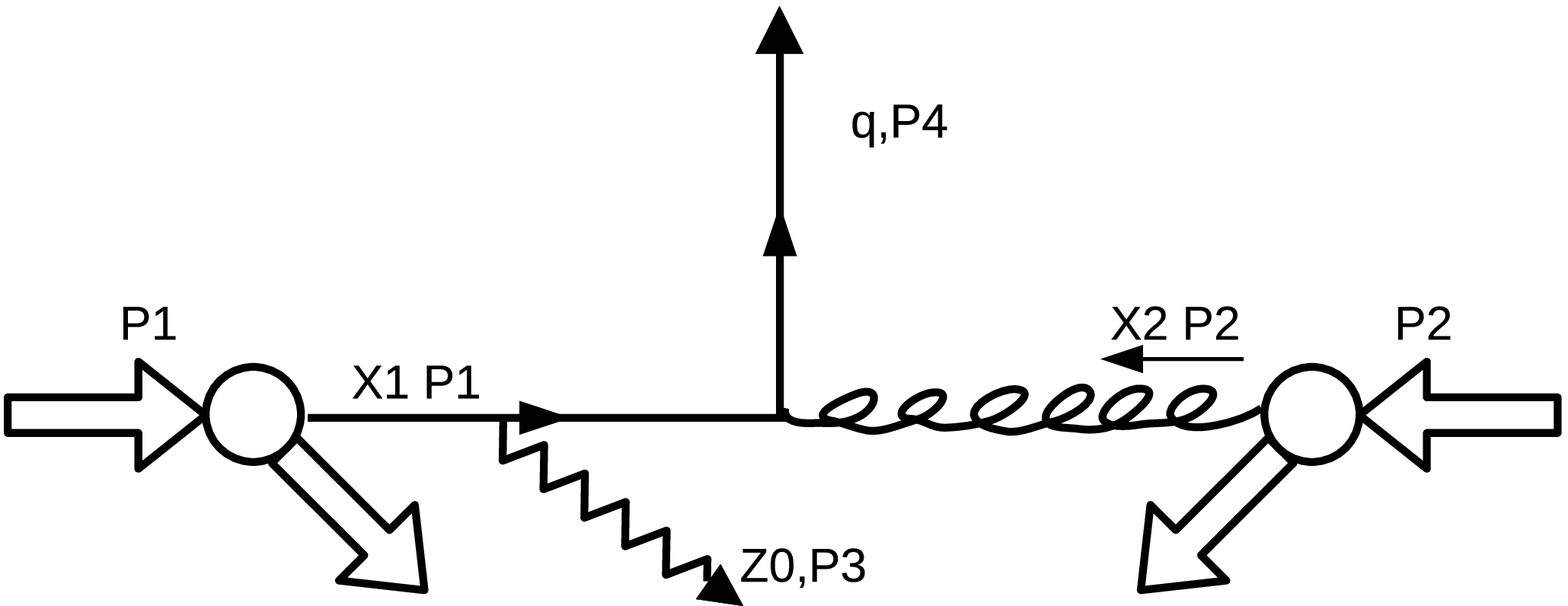}
\includegraphics[width = 0.4\linewidth]{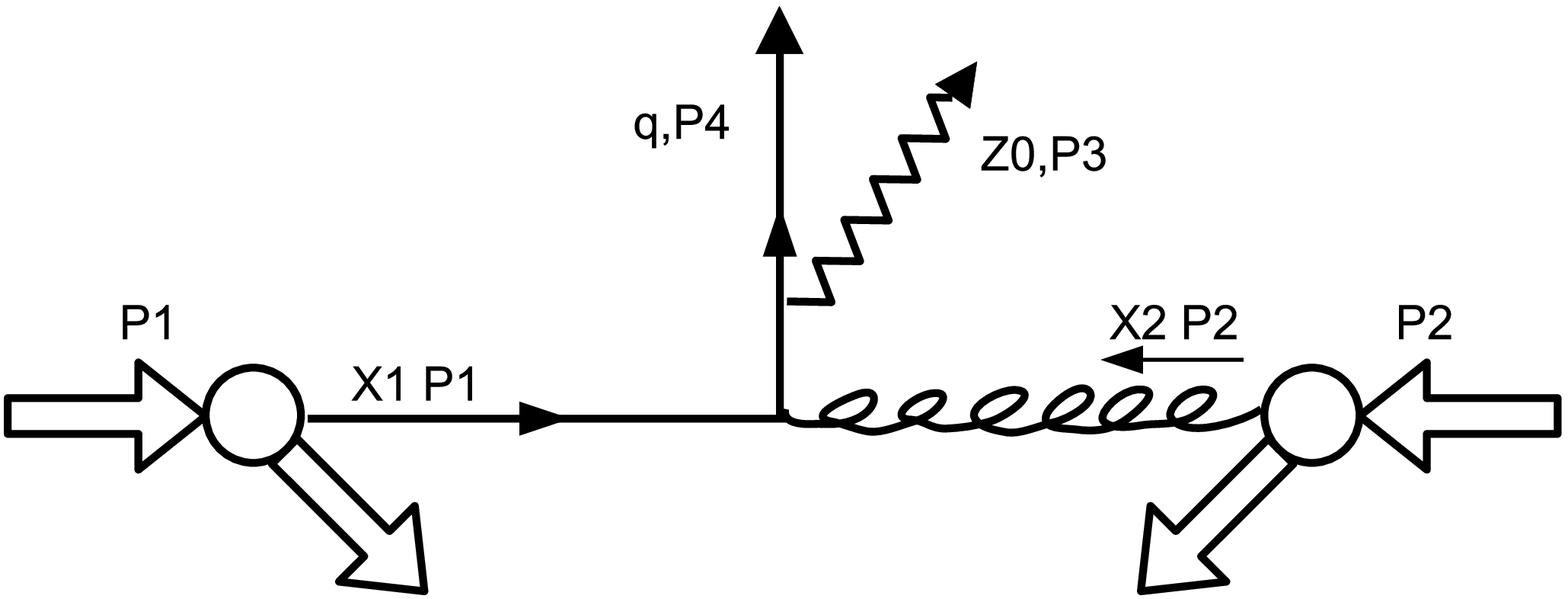}
}
\caption{\small{LO diagrams contributing to $Z^0$ tagged jet production.  Curly lines denote gluons, straight lines quarks, and jagged lines $Z^0$ (labeled with $Z0$ in the figure).  Two partons with momentum fraction $x_1$ and $x_2$ from nucleons with momenta $P_1$ and $P_2$, respectively, interact as seen above.  At LO the contribution from quark and gluon jets are conveniently separable; however, at NLO real gluon emission complicates that distinction.}}
\label{quark_glue}
\end{figure*}

Consider the cross section for $Z^0$ tagged jet production in $p+p$ collisions.  The leading-order (LO) production amplitude is represented by the Feynman diagrams in Figure \ref{quark_glue}.  In each case, the $Z^0$ is produced either from quark radiation or quark/anti-quark annihilation.  If one makes the approximation that the initial partons are collinear with the nucleon momenta, it is immediately clear that conservation of momentum requires the transverse (to the beam line) momentum (denoted $\bald{p}_T$ or $p_T = |\bald{p}_T|$ from here on) of the $Z^0$ and the scattered parton be equal.  At mid-rapidity, the $p_T$ of the associated $Z^0$ boson completely specifies the initial energy of the tagged jet at LO.

As mentioned above, I am here interested in the leptonic decay products of the $Z^0$ boson are.  For this reason, one must also include the contribution of virtual photon tagged jets in the invariant mass range of the $Z^0$.  Diagrammatically, this corresponds to making the replacement $Z^0\leftrightarrow \gamma^*$ in Figure \ref{quark_glue}.  In the results presented below, dilepton pairs in the invariant mass range $65-115$ GeV are considered (recall that $M_z = 91.1876$ GeV and $\Gamma_z = 2.4952$ GeV \cite{pdg}).  In this invariant mass range, the contribution from the $Z^0$ dominates that of the virtual photon by roughly two orders of magnitude.

Obtaining the LO result requires a lengthy but textbook level calculation, and the details will not be presented here. The LO result, in the notation of Field \cite{Field:1989uq}, for the matrix elements squared is
\begin{eqnarray*}
|M_g|^2 &=& \frac{16  \pi^2 \, \alpha_s \, \alpha\,(R_q^2 + L_q^2) }{9 x_w(1-x_w^2)} \left(\frac{\hat{t}^2 + \hat{u}^2 + 2\,\hat{s}\,m_z^2}{\hat{t} \hat{u}}\right) \\
|M_q|^2 &=& - \frac{6  \pi^2 \, \alpha_s \, \alpha\,(R_q^2 + L_q^2) }{9 x_w(1-x_w^2)} \left(\frac{\hat{s}^2 + \hat{t}^2 + 2 \, \hat{u} \,m_z^2}{\hat{s}\,\hat{t}}\right).
\end{eqnarray*}
When the initial-state partons are interchanged the matrix elements can be trivially obtained with the substitution $\hat{t} \leftrightarrow \hat{u}$.  In the above result, the subscript $q$ or $g$ refers to a quark or gluon jet, respectively, $\alpha_s = g_s^2/4\pi$, and the Lorentz-invariant Mandelstaam variables are defined such that
\begin{eqnarray*}
\hat{s} &=& (p_1 + p_2)^2  \\
\hat{t} &=& (p_1 - p_3)^2 \\
\hat{u} &=& (p_2 - p_3)^2
\end{eqnarray*}
where the notation of Figure \ref{quark_glue} is used.  Also, $x_w \equiv \sin^2\theta_w \approx 0.231$, $\alpha \approx 1/137.04$ \cite{pdg} and
\begin{eqnarray*}
R_q^2 &=& 4 \, e_q^2 \, \sin^4 \theta_w ,\\
L_q^2 &=& \tau_q^2 - 4 \, e_q \, \tau_q \, \sin^2 \theta_w + 4 \, e_q^2 \, \sin^4 \theta_w
\end{eqnarray*}
where $\tau_q$ is the weak isospin of quark $q$ (that is, $\tau = 1$ for $u,c,t$ and $\tau = -1$ for $d,s,b$) and $e_q$ is the fractional electric charge of quark $q$.  Modifying the above result to $\gamma^*$ tagged jets can be done with the replacement
\eqf{
\frac{(R_q^2 + L_q^2)}{x_w(1-x_w^2)}\rightarrow 2 e_q^2
}
as well as letting $m_z^2 \rightarrow Q^2$, where $Q$ is the photon mass.  The total scattering probability requires summing $|M_g|^2 + |M_q|^2$ over quark flavors.

The leptonic final states from the LO $Z^0/\gamma^*$ tagged jet cross section can be obtained by performing a Lorentz boost into the rest frame of the tagging particle.  In the rest frame of the particle, a large number of isotropically distributed random decays are performed, which are then boosted back into the lab frame.  The fraction of lepton pairs that are inside of the experimental limitations is then treated as a multiplicative factor, which, along with the appropriate branching ratio, is multiplied with the cross section.

\begin{figure}
\centerline{
\includegraphics[width = 0.5\linewidth]{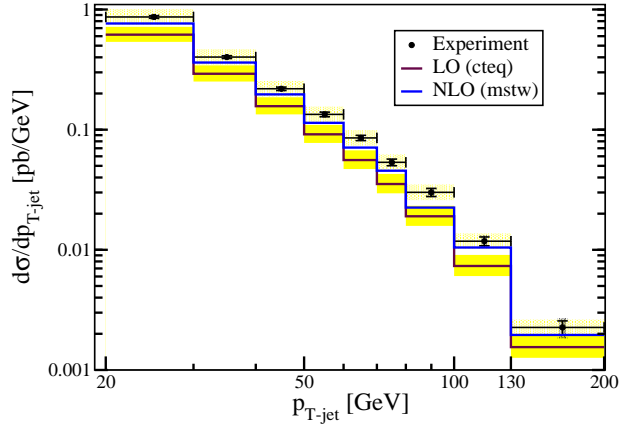}
}
\caption{\small{ Experimental results from the FermiLab Tevatron Collider \cite{Abazov:2008ez} of jets associated with $Z^0$/$\gamma^*\rightarrow \mu^+\mu^-$ in $p\bar{p}$ collisions at $\sqrt{s} = 1.96$ TeV.  A comparison of the MCFM NLO result (blue line) and the LO result (within the yellow bands) is included.  The experimental parameters are discussed in the text.  The (yellow) bands around the LO result indicated uncertainty in the choice of scale.}}
\label{dangy}
\end{figure}

Having the LO result is instructive and useful for understanding isospin effects when going from $p+p\rightarrow A+A$ or in obtaining the relative fraction of gluon versus quark jets.  However, in order to study the suppression due to medium induced energy loss, it is necessary to have the next-to-leading-order (NLO) result.  This is because, as mentioned above, at LO the $p_T$ of the $Z^0$ and the scattered parton are exactly equal.  Obviously medium induced energy loss in $A+A$ collisions will break this equality.  In order to make a meaningful comparison for the case when the $p_T$ of the $Z^0$ and the scattered parton are not equal (as is certainly the case in $A+A$), one must have the NLO result.  At NLO, the $p_T$ equality is smeared by real gluon emission, enabling a comparison between $p+p$ and $A+A$.

The NLO cross section is here obtained using the publicly available Monte Carlo for FeMtobarn processes (MCFM) developed by Campbell and Ellis \cite{Campbell:2002tg}, (available online at http://mcfm.fnal.gov/).  MCFM provides NLO predictions for many processes at hadron colliders.  The program has built-in options for a wide range of experimental acceptance cuts and run-time parameters.  A comparison of the LO result and the MCFM NLO result for jets associated with $Z^0$/$\gamma^*\rightarrow \mu^+\mu^-$ in $p+\bar{p}$ collisions at $\sqrt{s} = 1.96$ TeV with experimental measurements at the FermiLab Tevatron Collider \cite{Abazov:2008ez} is shown in Figure \ref{dangy}. The leading jet particle was required to have $p_T > 20$ GeV and rapidity $|y| < 2.8$, where
\eqf{
y = \frac{1}{2}\ln \frac{E-p_z}{E+p_z}
}
and $\hat{z}$ is oriented along the beam axis.  Jets were reconstructed using a midpoint cone algorithm with Lorentz-invariant cone size $R = \sqrt{(\Delta \phi)^2 + (\Delta y)^2} = 0.5$.  Additionally, the muons were required to have $p_T>15$ GeV and $|y| < 1.7$.  The yellow band around the LO result indicates the uncertainty due to the choice of scale in the problem.  Notice that the results shown in Figure \ref{dangy} are inclusive in the sense that the tagging particles are not constrained to lie within a certain $p_T$ interval.

\begin{figure*}
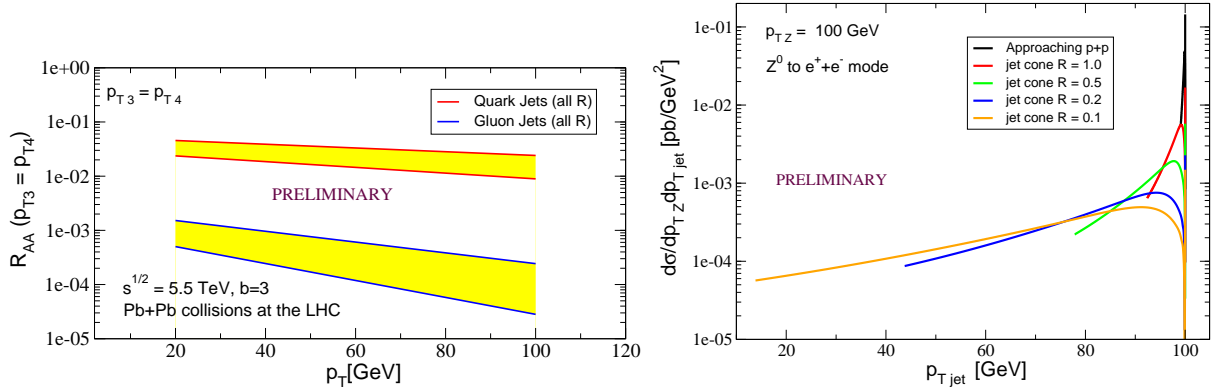

\centerline{
\includegraphics[width = 0.52\linewidth]{ExclussiveTag.eps}
\includegraphics[width = 0.46\linewidth]{ExclusDiff.eps}
}
\caption{\small{In a probabilistic description of partonic energy loss, the probability not to radiate, $P(0)$, can be directly related to the average number of gluons in the medium induced bremsstrahlung spectrum as a function of the $p_T$ of the parent parton.  This is the case considered in the left panel of the figure, $\ln R_{aa} = - \langle N_g \rangle$ (see discussion in text and equation (\ref{gn})).  In the right panel is plotted the modified nuclear cross section for different values of $R$ as a function of jet $p_T$.  As the jet cone radius, $R$, is increased the cross section approaches a $\delta$ function at $p_T = 100$ GeV, which is the $p_T$ of the tagging particle. At NLO, even as $R\rightarrow \infty$, the cross section would still have some broadening.  As $R$ is decreased, the result smears out, reflecting the fact that less and less of the initial energy is recovered in the jet.}}
\label{dongy}
\end{figure*}

The real advantage of using tagged jets is that one can require the tagging particles to lie in a certain $p_T$ range.  Results for the suppression of tagged jets at NLO will not be presented in this proceedings, as they are still in progress and will be forthcoming soon.  However, one can still obtain some information with LO $Z^0$/$\gamma^*\rightarrow \mu^+\mu^-$ tagged jets in $p+p$.  For instance, in a probabilistic description of partonic energy loss, the probability for a parton to lose some fraction of energy $\epsilon$ is described by the normalized distribution $P(\epsilon)$.  The probability for a parton not to lose energy, $P(0)$, is related to the average number of gluons emitted, $\langle N_g \rangle$.  In this specific case one has \cite{Vitev:2005he}
\eqf{\label{gn}
\ln R_{aa} = - \langle N_g \rangle.
}
This is precisely what is plotted in the left panel of Figure \ref{dongy} for central $Pb+Pb$ collisions $\sqrt{S} = 5.5$ TeV per nucleon.  The results were obtained using the GLV partonic energy loss formalism.  From the Figure, one can determine the expected number of gluons emitted as a function of the $p_T$ of the parent parton in this formalism.

One can also examine the sensitivity of the jet cross section in $Pb+Pb$ on cone radius.  Again, using the LO result for $Z^0$/$\gamma^*\rightarrow \mu^+\mu^-$ tagged jets in $p+p$ and the GLV formalism, the modified nuclear cross section is plotted for different values of $R$ as a function of jet $p_T$ in the right panel of Figure \ref{dongy}.  As the jet cone radius, $R$, is increased the cross section approaches a $\delta$ function at $p_T = 100$ GeV, which is the $p_T$ of the tagging particle. At NLO, even as $R\rightarrow \infty$, the cross section would still have some broadening.  As $R$ is decreased, the result smears out, reflecting the fact that less and less of the initial energy is recovered in the jet.

In summary, I have argued that leading particle suppression alone is not sufficient to discriminate between partonic energy loss formalisms or to extract quantitatively the jet quenching properties of the QGP.  New and more differential observables are needed to do this.  In particular, jet shapes and jets tagged with weakly interacting probes are especially promising.  The heavy-ion program at the LHC will enable high statistics experimental measurement of these observable for the first time.  It is expected to usher in a new era in the study of QGP induced partonic energy loss and other related areas of QGP physics, such as medium response \cite{Neufeld:2008fi,Neufeld:2009ep} in the context of jets tagged with direct and virtual photons and electroweak bosons.  I have here presented preliminary results for the production and nuclear modification of jets associated with $Z^0$/$\gamma^*$ bosons at LHC energies using the GLV partonic energy loss formalism.  The full results of this study are forthcoming.

$$$$

\small{{\it Acknowledgments}: I wish to thank my collaborators, Ivan Vitev and Ben-Wei Zhang, who have contributed equally in this work.  This work was supported in part by the US Department of Energy, Office of Science, under Contract No. DE-AC52-06NA25396.}

$$$$

 % do not change
\end{document}